\documentclass{aa}
\usepackage{graphicx}
\usepackage{txfonts}
\begin{document}
   \title{A Dark Energy model combining DGP gravity and Chaplygin gas}

   \author{M. Roos}
   \offprints{M. Roos}

   \institute{Department of Physical Sciences and Department of
          Astronomy, FIN-00014 UNIVERSITY OF HELSINKI\\
              \email{matts dot roos at helsinki dot fi}}
              \date{Received April 2007; accepted }


  \abstract
   {The expansion of the Universe is
accelerating, as testified by observations of supernovae of type Ia
as a function of redshift. Explanations of this acceleration are of
two types: modifications of Einstein gravity or new forms of energy,
coined dark energy.}
   {The accelerated expansion is explained here by
a combination of Dvali-Gabadadze-Porrati (DGP) model gravity and
Chaplygin gas dark energy. Both models are characterized by a length
scale L which may be the same.}
   {The continuity equation for the combined model is derived in flat
   geometry, and solved by numerical methods.}
   {The solution is shown to have the expected
properties: at very small scales ($a \ll L$) the energy density
behaves as pressureless dust, at very large scales ($a \gg L$) as a
cosmological constant. The modifications to the DGP model and the
Chaplygin gas model occur for values of $a$ near $L$.} {The results
show an increase in the present dark energy density relative to the
plain DGP model.}


   \maketitle

\section{Introduction}

The demonstration by SNeIa that the Universe is undergoing an
accelerated expansion has stimulated a vigorous search of models to
explain this unexpected fact. Since the dynamics of the Universe is
conventionally described by the Friedmann equations which follow
from the Einstein equation in four dimensions, all departures from
the $\Lambda CDM$ concordance model must be due to some modifications
of the Einstein equation.

The left-hand-side of the Einstein equation encodes the geometry of
the Universe in the Einstein tensor $G_{\mu\nu}$, the
right-hand-side encodes the energy content in the stress-energy
tensor $T_{\mu\nu}$. Thus modifications to $G_{\mu\nu}$ imply some
alternative gravitation, modifications in $T_{\mu\nu}$ involve new
forms of energy densities that have not been observed, and which
therefore are called dark energy.

A well-studied model of modified gravity is the
Dvali-Gabadadze-Porrati (DGP) braneworld model (Dvali \& al. \cite{Dvali}) in which our 4-dimensional world is a FRW brane embedded in a
5-dimensional Minkowski bulk. On the 4-dimensional brane the action
of gravity is proportional to $M^2_{Pl}$ whereas in the bulk it is
proportional to the corresponding quantity in 5 dimensions, $M^3_5$.
The model is then characterized by a cross-over length scale
 \begin{equation}
L=\frac{M^2_{Pl}}{2M^3_5}\ ,\label{L}
 \end{equation}
such that gravity is a 4-dimensional theory at scales $a \ll L$
where matter behaves as pressureless dust, but gravity "leaks out"
into the bulk at scales $a \gg L$ and matter approaches the
behaviour of a cosmological constant. To explain the accelerated
expansion which is of recent date ($z\approx 0.5$ or $a\approx 2/3$), L must be of the order of 1.

An equally interesting model introduces a fluid called Chaplygin gas
(Kamenshchik \& al. \cite{Kamenshchik}, Bilic \& al. \cite{Bilic}) following work in aerodynamics (Chaplygin \cite{Chaplygin}). This addition to $T_{\mu\nu}$ is intriguingly
similar to the DGP model in the sense that it is also characterized
by a cross-over length scale below which the gas behaves as
pressureless dust, and above which it approaches the behaviour of a
cosmological constant. This length scale is expected to be of the
same order of magnitude as the L scale in the DGP model.

Both the DGP model and Chaplygin gas have problems with
fitting present cosmological data (briefly summarized by Copeland \& al. \cite{Copeland}). In a comparison with supernova data the DGP model is slightly disfavored, at the level of $2\sigma$ (Rydbeck \& al. \cite{Rydbeck}). In the Chaplygin gas model the Jeans instability of perturbations behaves like CDM fluctuations in the dust-dominated stage ($a\ll L$), but disappears in the acceleration stage ($a\gg L$). The combined effect of suppression of perturbations and non-zero Jeans length leads to a strong ISW effect and thus of loss of power in CMB anisotropies (Amendola \& al. \cite{Amendola}, Bento \& al. \cite{Bento}).

This has led to generalizations which are less motivated than the original models. Below we shall discuss a model which combines both models, motivated by the
similarities in their asymptotic properties and in the cross-over scales.

\section{Continuity equations}

The Friedmann equation for Hubble expansion in the DGP model may be written (Deffayet \& al. \cite{Deffayet})
\begin{equation}
H^2-\frac k{a^2}\pm\frac 1 L\sqrt{H^2-\frac k{a^2}}=\kappa\rho~,\label{F1}
\end{equation}
where $\kappa=8\pi G/3$, and $\rho$ is the total cosmic fluid energy density. In the following we shall set $k=0$ corresponding to a flat geometry. A + sign in front of the root term causes accelerating expansion.

The Friedmann equation for the rate of change in $H$ is simplest written
\begin{equation}
\frac{2\ddot a}a+H^2=-3\kappa p~,\label{F2}
\end{equation}
where $p$ is the total cosmic fluid pressure. Pressureless dust has $p=0$, in the $\Lambda CDM$ model the pressure is $p_{\Lambda}=-\rho_{\Lambda}$, in the Chaplygin gas model $p_{\varphi}=-A/\rho_{\varphi}$, where $A$ is a constant. Differentiating Eq.~(\ref{F1}) and using it to eliminate the second time derivative in Eq.~(\ref{F2}) one obtains the continuity equation (or energy conservation equation).

In the ordinary FRW universe, the continuity equation for Chaplygin gas is
\begin{equation}
\dot\rho_{\varphi}(a)-3H\left(\rho_{\varphi}(a)-\frac{A}{\rho_{\varphi}(a)} \right)=0~.
\end{equation}

In the DGP universe the continuity equation for a generic energy density $\rho(a)$ with pressure $p(a)$ is
\begin{equation}
\dot\rho(a)=-3\frac{\dot a}{a} \left(\rho(a) + p(a) + \frac{1}{2\kappa L^2} \right)+ \frac{3}{2L}[\rho(a)-p(a)].\label{eq1}
\end{equation}
Since ordinary matter does not interact with Chaplygin gas, one can derive separate continuity equations for the energy densities $\rho_m$ and $\rho_{\varphi}$, respectively. Thus ordinary matter obeys Eq.~(\ref{eq1}) (5) with $p=0$, our combined model for Chaplygin gas in a DGP universe obeys
\begin{equation}
\dot\rho_{\varphi}(a)=-3\frac{\dot a}{a} \left(\rho_{\varphi}(a)-\frac{A}{\rho_{\varphi}(a)} + \frac{1}{2\kappa L^2} \right)+ \frac{3}{2L}\left(\rho_{\varphi}(a)+\frac{A}{\rho_{\varphi}(a)}\right). \label{eq2}
\end{equation}

\section{Solutions}

The continuity equation for Chaplygin gas integrates to
\begin{equation}
\rho_{\varphi}(a)=\sqrt{A+\frac B {a^6}}~,\label{C}
\end{equation}
where $B$ is an integration constant. Thus this models has two free parameters. Obviously its limiting behaviour is
\begin{equation}
\rho_{\varphi}(a)\propto\frac{\sqrt{B}}{a^{3}}~~ {\rm for}~~ a~\ll \left(\frac B A\right)^{1/6},~~~\rho_{\varphi}(a)\propto -p~~  {\rm for}~~ a\gg \left(\frac B A\right)^{1/6}.
\end{equation}
The identification of the cross-over limit here with L in the DGP model implies $B/A\approx L^6$. However, this cannot be substituted into the combined model, because the latter may have a different integration constant. Moreover, Eq.~(\ref{eq2}) can only be integrated numerically, and then no value can be substituted into it.

\begin{figure*}
   \includegraphics[width=9cm]{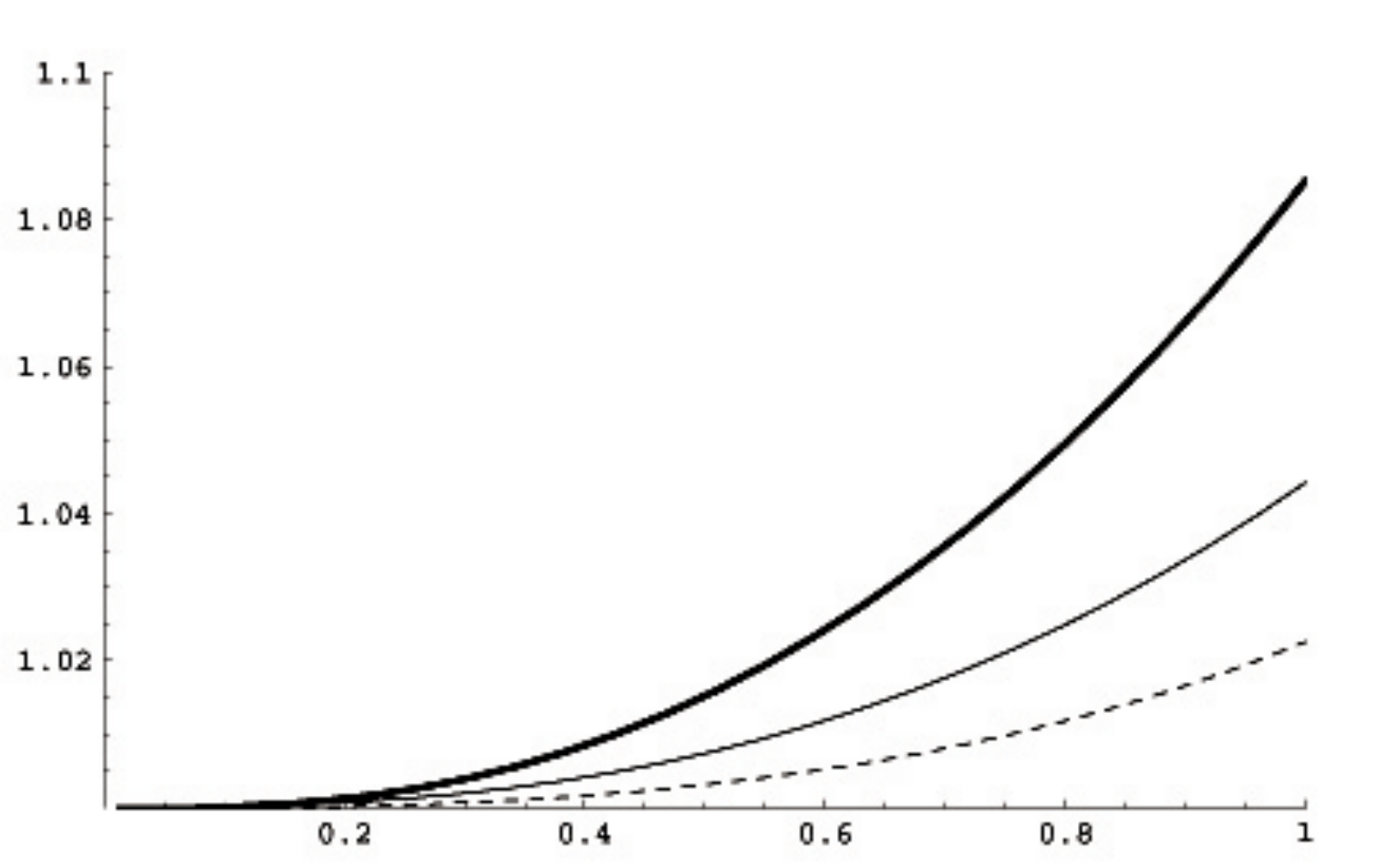}
   \caption{The ratio $R(a,L)$ of the density $\rho_{\varphi}(a)$ of the combined model\newline to the DGP density $\rho_m(a)$ for $L=1.0$ (thick solid line), $L=1.2$ (thin\newline solid line), and $L=1.4$ (dashed line).
$L=1$ corresponds to $a=0.5$.}
    \end{figure*}

In the case of pressureless dust the integration of Eq.~(\ref{eq1}) with $p=0$ gives
\begin{equation}
\rho_m(a)=a^{-3}\left(\frac{8\kappa L}{9}+\frac{4\kappa a}{3}+\frac{\kappa a^2}{L}+C \exp\left(\frac{3a}{2L}\right)\right)~,\label{rhom}
\end{equation}
where $C$ is an integration constant. Otherwise there is only one free parameter, $L$.

Obviously $\rho_m(a)\propto a^{-3}$ for small values of $a/L$. The density must also be positive for all $a$, thus $C$ must be non-negative. One can fix the value of $C$ by some boundary condition at $a\gg L$ without affecting the $a$-dependence of $\rho_m(a)$ noticeably in the cosmologically interesting region $a\leq 1$. For $a/L> 1$, $\rho_m(a)$ approaches an $a^{-1}$ quintessence-like behaviour or de Sitter-like behaviour. Ultimately the exponential term in Eq.~(\ref{rhom}) begins to dominate, leading to a phantom-like solution. By choosing the integration constant $C$ sufficiently small one can shift the de Sitter behavior and subsequent phantom behavior into the distant future without affecting the shape of $\rho_m(a)$ in the region $a\leq 1$.

The continuity equation (\ref{eq2}) cannot be integrated analytically, but only numerically. Then a boundary condition for $\rho_{\varphi}(a)$ needs to be specified. In addition the solution depends on the two parameters $A$ and $L$. The properties of the solution can best be studied graphically.

The function $a^3\rho_{\varphi}(a)$ approaches a constant for $a\rightarrow 0$, thus the $a$-dependence of $\rho_{\varphi}(a)$ is $a^{-3}$ for all choices of $A$ and $L$, in agreement with the functions (\ref{C}) and (\ref{rhom}). To permit a comparison of $\rho_{\varphi}(a)$ and $\rho_m(a)$, the latter is normalized to unity at $a=0$, and $a^3\rho_{\varphi}(a)$ is scaled to unity with an arbitrary constant at $a=0.01$ (because the numerical function is singular at $a=0$).

For $a>L$ the $a$-dependence of $\rho_{\varphi}(a)$ slows down, approaching a "cosmological" constant. At some later point the solution starts to grow exponentially in a phantom-like behavior, just as was the case for $\rho_m(a)$ in Eq.~(\ref{rhom}). This cannot be cured exactly because of the numerical nature of the solution to Eq.~(\ref{eq2}), but the problem can be shifted into the distant future by a proper choice of boundary condition. The exact position and function value at the point of the boundary condition only affects the normalization of $a^3\rho_{\varphi}(a)$, but not its shape in the cosmologically interesting region $a\leq 1$.

To appreciate the changes brought to the DGP model by Eq.~(\ref{eq2}) we shall study the ratio
\begin{equation}
R(a,L)\equiv \frac{\rho_{\varphi}(a,L,A)}{\rho_m(a,L,C)}.\label{R}
\end{equation}
Ideally, the boundary condition in Eq.~(\ref{eq2}) should be chosen so that the magnitude of the exponential term hidden in the numerical solution $\rho_{\varphi}(a)$ would exactly compensate $C$ in Eq.~(\ref{rhom}), in which case $R(a,L)$ would approach a constant value when $a\gg L$. However, to meet such a condition is not possible with a numerical solution, thus one has to resort to represent the effect of both integration constants by a normalization factor, here chosen to give $R(0.01,L)=1$.

The interesting results of this model are the $A$- and $L$- dependences in the region of observed acceleration $0.5<a<1$ (or $1>z>0$). We have found that $A$ acts as a mere factor, that can be considered as included in the normalization condition for the ratio (\ref{R}). (For this reason we have not defined $R$ to be a function of A.)

Figure 1 gives an example of the $L$- dependence: we plot the ratio $R(a,L)$ in the range $0.01\leq a\leq 1$ for $L=1.0$ (thick solid line), $L=1.2$ (thin solid line), and $L=1.4$ (dashed line). The $L$-scale has been chosen as the double of the $a$-scale: $L=1$ corresponds to $a=0.5$ or $z=1$. It is clear from the Figure that $\rho_{\varphi}(a)$ entails corrections to the DGP-model density $\rho_m(a)$ of the order of several percent in the region $a\approx 0.5-1$, where accelerated expansion has been observed.

\section{Conclusions}
We have studied a model of dark energy in which the geometry of the Universe is described by a brane in a 5-dimensional bulk: the DGP model, and where Chaplygin gas enters as a component in the stress-energy tensor.

This combined model depends on two parameters, the DGP model cross-over length $L$ and the Chaplygin gas parameter $A$, the latter entering only as a normalization parameter. In the region $a\approx 0.5-1$ where accelerated expansion has been observed, the model density exceeds corrections to the DGP density by several percent, as can be seen in Fig.~1. Correspondingly, the density parameter $\Omega_{\varphi}$ increases at fixed $\Omega_m$.

In this letter no fit to SNeIa data has yet been done. CMB anisotropy data can only be used when power spectra have been derived for the 5-dimensional bulk space implied here. It remains to be studied whether the various shortcomings of the DGP model and the Chaplygin gas model then are overcome.

\end{document}